\documentstyle[12pt]{article}
\font\twlmsy=msym10 at 12pt
\font\sevenmsy=msym8
\font\fivemsy=msym6
\newfam\Bbbfam
\textfont\Bbbfam=\twlmsy
\scriptfont\Bbbfam=\sevenmsy
\scriptscriptfont\Bbbfam=\fivemsy

\newcommand{\rf}[1]{(\ref{#1})}

\newcommand{\beq}{\begin{equation}}
\newcommand{\eeq}{\end{equation}}
\newcommand{\bea}{\begin{eqnarray}}
\newcommand{\eea}{\end{eqnarray}}
\newcommand{\beas}{\begin{eqnarray*}}
\newcommand{\eeas}{\end{eqnarray*}}
\newcommand{\beqs}{\begin{displaymath}}
\newcommand{\eeqs}{\end{displaymath}}

\newcommand{\cL}{{\cal L}}

\newcommand{\ben}{\begin{equation}}
\newcommand{\een}{\end{equation}}

\newcommand{\bdm}{\begin{displaymath}}
\newcommand{\edm}{\end{displaymath}}

\newcommand{\pa}{\partial}

\newcommand{\cN}{{\cal N}}

\begin{document}
\topmargin 0pt
\oddsidemargin 5mm
\headheight 0pt
\topskip 0mm

\addtolength{\baselineskip}{0.20\baselineskip}

\pagestyle{empty}

\hfill October 1994

\bigskip

\begin{center}

\vspace{12pt}
{\Large \bf
Remarks on The Entropy of 3-Manifolds}

\vspace{1 truecm}

\centerline{}

\vspace{2 truecm}

{\em Bergfinnur Durhuus\footnote{e-mail: durhuus@math.ku.dk}}

\medskip

Matematisk Institut, University of Copenhagen \\
Universitetsparken 5, 2100 Copenhagen \O \\
Denmark

\vspace{1.3 truecm}

{\em Thordur Jonsson\footnote{e-mail: thjons@raunvis.hi.is}}

\medskip

Raunvisindastofnun Haskolans, University of Iceland \\
Dunhaga 3, 107 Reykjavik \\
Iceland

\vspace{2 truecm}

\end{center}

\noindent
{\bf Abstract.}
 We give a simple combinatoric proof of an exponential
upper bound on the number of
distinct 3-manifolds that can be
constructed by successively identifying  nearest
neighbour pairs of
triangles in the boundary of
a simplicial 3-ball and show that all closed simplicial
manifolds that can be constructed in this manner are homeomorphic to $S^3$.
We discuss the problem of proving that all 3-dimensional
simplicial spheres can be obtained by this construction
and give an example of a simplicial 3-ball whose boundary
triangles can be identified pairwise such that no triangle is
identified with any of its neighbours and the resulting
3-dimensional simplicial complex is a simply connected 3-manifold.

\vfill

\newpage
\pagestyle{plain}

\section{Introduction}
A few years ago a model based on random simplicial 3-manifolds was
introduced as a discretization of 3-dimensional quantum gravity \cite{adj1}
in the same spirit as randomly triangulated surfaces have been used to
study quantum gravity in 2 dimensions \cite{rs1,rs2,rs3}.
This model has been discussed by several authors,
generalized to higher dimensions and
extensively simulated, see e.g. \cite{migdal1,numamb,numcat,nummig,a,b,c,d}.

It has not yet been proven that these models have a convergent grand
canonical partition function.  The problem
is to prove that the number $\cN (N)$ of combinatorically
distinct, but topologically identical, simplicial manifolds that can be
constructed using $N$ tetrahedra (or $d$-simplexes in $d$ dimensions)
satisfies an exponential bound
\beq
\cN (N)\leq C^N\label{bound}
\eeq
for some constant $C$.
Numerical simulations indicate though
that the bound holds, at least in 3
dimensions  \cite{x,y}.  In \cite{adj1} some sufficient
technical conditions for the exisitence of the desired bound for $d=3$ were
discussed.

Considering the general interest in and importance of this question, see
\cite{y,cattclaim,janrebuttal,boulatov}, we find it worthwhile to report
on some partial results towards proving \rf{bound} and
point out the principal obstacle to completing the argument.  In particular,
we show that a claimed proof \cite{boulatov} of \rf{bound} is based on a
false assumption about the structure of simplicial 3-manifolds to which
we provide a counterexample.  For the 3-manifolds that
satisfy this assumption we give a simple proof of an exponential bound
for $\cN (N)$.


\section{An exponential bound}
Any closed simplicial 3-manifold can be constructed by taking a suitable
simplicial ball and identifying the triangles in the boundary pairwise.
The resulting simplicial complex is a manifold if and only if its Euler
characteristic is 0 \cite{seifert}.  The Euler characteristic of a
3-dimensional simplicial complex $M$  is defined as
\beq
\chi (M)=\sum_{i=0}^3 (-1)^iN_i(M)
\eeq
where $N_i(M)$ is the number of $i$-simplexes in the complex.

It follows from this that one can construct any simplicial 3-manifold by first
taking a tetrahedron, gluing on it another tetrahedron
along a common triangle
and so on until
there are no more tetrahedra to be added.  In this way one obtains a
simplicial ball.  Then one identifies pairwise the
triangles in the boundary of the resulting simplicial ball.  It is not
hard to prove, and we shall give the argument below,
that the number of
distinct simplicial balls that can arise in the first step of
this construction increases
at most exponentially with the number of tetrahedra.  In general the
number of triangles in the boundary of the simplicial ball is of the
order of the number of tetrahedra and the possibility of
superexponential factors arises when these triangles are identified
pairwise.
If one does not put any restriction on the topology of the resulting
manifold the number of distinct manifolds does indeed grow
factorially with the number of tetrahedra \cite{adj1}.

The Euler characteristic of a three-dimensional simplicial ball is $1$.
Suppose we are given such a ball $B$ and a pairwise  identification of its
boundary triangles so that we obtain a simplicial complex  $M$ after the
identifications have been carried out.  After
the identifications the triangles that were originally
in the boundary of $B$ form a two-dimensional
complex $K$ inside $M$.
It is easily seen that the two-dimensional Euler characteristic, $\chi
_2 (K)$, of
the complex $K$, defined as
\beq
\chi _2(K)=\sum _{i=0}^{2}(-1)^iN_i(K),
\eeq
equals $1+\chi (M)$ so $M$ is a manifold if and only if $\chi _2(K)=1$.
Moreover, the first homotopy group of $M$ is identical to the first
homotopy group of the complex $K$ \cite{seifert}.

Let us consider the problem of constructing a 3-manifold with the
topology of the three-dimensional sphere $S^3$.  Suppose we are given a
simplicial ball and we want to identify the triangles in its surface
pairwise so that we get a manifold homeomorphic to $S^3$.  One of the
conditions
that the identification of triangles must satisfy (and the only
condition if the Poincar\'{e} conjecture holds) is that
noncontractible loops should not arise in the resulting complex.  One
would like to exhibit a local condition on the identifications that
ensures this property and at the same time allows the construction of
all possible simplicial spheres.  For the analogous construction of
$S^2$ by gluing 1-simplexes together this problem is solved (see
\cite{erice}) by allowing only successive identifications of
neighbouring links, i.e. pairs of links sharing a vertex.  It is
therefore reasonable to conjecture that for $S^3$ one should require
that at each step
identification of two triangles is only allowed if they share a
link.  More precisely,
we say that a simplicial 3-manifold $M$ has a {\em
local construction} if there is a sequence of simplicial manifolds
$T_1,\ldots ,T_n$ such that
\begin{itemize}
\item[(i)] $T_1$ is a tetrahedron
\item[(ii)] $T_{i+1}$ is constructed from $T_i$ by either gluing a new
tetrahedron to $T_i$ along one of the triangles in the boundary $\pa
T_i$ of $T_i$ or by identifying a pair of nearest neighbour triangles in
$\pa T_i$, i.e. two triangles sharing a link in $\pa T_i$.
\item[(iii)] $T_n=M$
\end{itemize}
We shall
denote by $\cL$
the collection of all closed simplicial 3-manifolds that have a local
construction.  In the next section we show that all
manifolds in $\cL$ are homeomorphic to $S^3$.
It is, however, not known whether all simplicial spheres have a local
construction.   This was claimed for all simply connected 3-manifolds in
\cite{boulatov} but as we shall see below the argument given there is
not complete.

First we give a simple proof that the number of complexes in $\cL$ with
a given number $N$ of 3-simplexes is bounded by an exponential function
of $N$.

\medskip

\noindent
{\bf Theorem 1} {\it There is a constant $C$ such that}
\beq
\#\{ M\in\cL :N_3(M)=N\}\leq C^N.\label{bound2}
\eeq
{\bf Proof.} It is easy to see that if a manifold has a local
construction and contains $N$ tetrahedra it also has a local
construction where all the tetrahedra are assembled in the first $N$
steps.  In the first $N$ steps we therefore build a tree-like 3-manifold
with at most 4 branches emanating from each node.  In the dual picture
(where a tetrahedron is a vertex and a pair of identified triangles is a
link joining the corresponding vertices) such a 3-manifold  is a
$\varphi ^4$ tree-graph with $N$ vertices.  The number of distinct
graphs of that kind is well-known to be bounded by $C_1^N$ where $C_1$
is a constant.

Let us now assume that we have one of the tree-like simplicial
manifolds described
above made of $N$ tetrahedra.  Its surface consists of $2N+2$ triangles.
 We wish to estimate from above how many ways there are to close up this
manifold by successively identifying nearest neighbour pairs of
triangles.  Suppose that in the beginning there are $n_1$ pairs of
nearest neighbour triangles that are to be identified.  Once these
identifications have been carried out the remaining unidentified
triangles in general have new neighbours and there are $n_2$ pairs of
nearest neighbour triangles that are to be identified.  We continue in
this fashion, identifying $n_i$ pairs in the $i$-th
step until there are no triangles left after $f$ steps.  Clearly $f\leq
N+1$ because $n_i\geq 1$.

The number of ways to choose the $n_1$ pairs
of triangles that participate in the first round of
identifiactions is bounded by
\beq
\left(\begin{array}{c}2N+2\\ n_1\end{array}\right)3^{n_1}.
\eeq
After carrying out these identifications there arise at most $2n_1$ new
pairs of nearest neighbour triangles that might be identified, see Fig. 1.

In the next step of the construction we choose $n_2$ triangles out of
the $4n_1$ triangles that possibly may be identified with one of their
neighbours and identify each of them with one of their
neighbours.  The number of
ways this can be done is bounded by
\beq
\left(\begin{array}{c}4n_1\\ n_2\end{array}\right)3^{n_2}.
\eeq
We continue in this fashion until there are no triangles left.
Clearly
\beq
2\sum _{i=1}^fn_i=2N+2
\eeq
where $2n_f$ is the number of triangles left before the final step in
the identification process is carried out.

The total number of ways one can close the tree-like manifold is
therefore bounded by
\bea
& & \sum _{f=1}^{N+1}\sum _{n_1,\ldots ,n_f}
\left(\begin{array}{c}2N+2\\ n_1\end{array}\right)
\left(\begin{array}{c}4n_1\\ n_2\end{array}\right)
\left(\begin{array}{c}4n_2\\ n_3\end{array}\right)\ldots
\left(\begin{array}{c}4n_{f-1}\\ n_f\end{array}\right)3^{N+1}\nonumber\\
&\leq &\sum _{f=1}^{N+1}\left(\begin{array}{c}N\\
f-1\end{array}\right)
2^{6N+6}3^{N+1}  \nonumber\\
& \leq & \sum _{f=1}^{N+1}2^{7N+6}3^{N+1}\nonumber\\
& \leq & C_2^N.
\eea
This completes the proof with $C=C_1C_2$.

\medskip

Note that in the bound derived above we have not used that $\chi =0$ for
3-manifolds so we have in fact established a bound on the number of
pseudomanifolds with a local construction.  A bound of the form
\rf{bound2} is also obtained in \cite{boulatov} but with a more
elaborate proof.

\section{Properties of manifolds with a local construction}
We begin by demonstrating that the elements in $\cL$ are simplicial
spheres.  This is a consequence of the following result.

\medskip
\noindent
{\bf Theorem 2} {\it Let $T_1,\ldots ,T_n$ be a local construction of a
simplicial manifold $M$. Then, for all $i=1,\ldots ,n$, $T_i$ is
homeomorphic to $S^3$ with a number of simplicial 3-balls removed.  The
boundary $\pa T_i$ is a union of simplicial 2-spheres, $S_1,\ldots ,S_k$,
and these fulfill:
\begin{enumerate}
\item[(i)] $S_r$ and $S_s$ have at most one point (vertex) in common for
$r\neq s$, $1\leq r,s\leq k$
\item[(ii)] The connected components of $\pa T_i$ are simply connected.
\end{enumerate}
}
{\bf Proof.} This is a straightforward inductive argument.  The
single tetrahedron $T_1$ is
homeomorphic to a closed 3-ball which may be
regarded as $S^3$ with an open 3-ball removed.

Assume that $T_i$ satisfies the properties listed in the theorem.
If $T_{i+1}$ is obtained from $T_i$ by
gluing on a tetrahedron this clearly does not change the homeomorrphism class
of the manifold and $T_{i+1}$ has the properties listed in the theorem.
On the other hand, when gluing together two neighbouring
triangles in $\pa T_i$ to obtain $T_{i+1}$ a number of distinct possibilities
have to be considered.
Note, however, that in all cases the two triangles belong to the same
2-sphere $S_r\subset\pa T_i$ due to condition (i).

(a) The two triangles have only one link in common.  Thus, the two
different vertices $P_1$ and $P_2$ in $\pa T_i$ opposite to this link are
identified.   In this case the sphere $S_r$ shrinks to a new sphere
containing 2 fewer triangles.  Those of the other spheres that have $P_1$ or
$P_2$ as a contact point with $S_r$ now all have a common contact point
with the new sphere.  Because of (ii) this is the only point shared by
any two of the spheres involved and clearly the boundary remains
simply connected.

(b) The two triangles have one link and the point $P$ opposite to the
link in common.  In this case the sphere $S_r$ splits into two spheres
with one point in common, namely $P$.  It is easy to check
that conditions (i) and
(ii) still hold.

(c) The two triangles have two links in common and these two links
emerge from a vertex $P$.  In this case the boundary
spheres touching $S_r$ at
$P$ are split off when we identify the triangles
and $S_r$ shrinks to a new sphere. This gives rise to
at most one more connected component in the boundary of the manifold
and conditions (i) and (ii) are still satisfied.

(d) The two triangles have all three edges in common.  In this case
$S_r$ consists solely of these two triangles and $S_r$ disappears upon
their identification.  The boundary will in general split into three
parts but still satisfies (i) and (ii).

\medskip

As mentioned above we do not have a proof that every simplicial 3-sphere
has a local construction.  Given a simplicial 3-sphere $M$ made of $N$
tetrahedra a possible strategy to produce such a local construction is
first to assemble all the $N$ tetrahedra as described in the proof of
Theorem 1 in the previous
section and then successively identify nearest neighbour triangles.
Continuing in this fashion one ends up with a simplicial manifold $M'$
which has the properties described in Theorem 2 and no two neighbouring
triangles in $\pa M'$ may be identified in order to construct $M$.  One
must then show that the requirement that $M$ be homemorphic to $S^3$
implies that $M'=M$, i.e. $\pa M'=\emptyset$.  The following result is a
step in this direction.

\medskip
\noindent
{\bf Proposition 3} {\it Let $M'$ be any simplicial 3-manifold with a nonempty
boundary having the properties described in Theorem 2.  Let $M$ be a
closed 3-manifold constructed from $M'$ by identifying the triangles in
$\pa M'$ pairwise in such a way that at least two triangles from
different boundary spheres $S_r$ and $S_s$ are identified.  Then $M$ is not
homeomorphic to $S^3$. }

\smallskip
\noindent
{\bf Proof.} Let $S_r$ be a 2-sphere in the boundary of $M'$ and choose
a collar neighbourhood $C_r$ of $S_r$ which may be pinched at the
finite number of points where
$S_r$ meets other boundary spheres.  The boundary of $C_r$ consists of
$S_r$ and another 2-sphere $\bar{S}_r$ which lies in the interior of $M'$
except at the pinching points which it shares with $S_r$.  Suppose now
that a triangle $t$ in $S_r$ is identified with another triangle $t'$ in
$S_s$, $r\neq s$.  Consider a smooth curve connecting a point $p$ in $t$
with the corresponding point $p'$ in $t'$.  We can assume that the curve
intersects $\bar{S}_r$ exactly once.  In $M$ this curve will be closed
since the endpoints have been identified, and it intersects $\bar{S}_r$
once.  If $M$ is homeomorphic to $S^3$ this contradicts the
Jordan-Brouwer theorem \cite{spivak}, which states that any 2-sphere in
$S^3$ separates it into two connected components and hence any smooth closed
curve which intersects the 2-sphere transversally must intersect it an
even number of times.  The fact that in the present case the sphere may
be pinched at a finite number of points does not affect the argument.

\medskip
We remark that the proof above uses explicitly that that $M$ is
homeomorphic to $S^3$ and not only that $M$ is simply connected.
Proposition 3 implies that in $\pa M'$ only identifications of triangles
within the same 2-sphere in the boundary are allowed if the resulting
closed manifold $M$ is homeomorphic to $S^3$.  In order to establish the
existence of a local construction for all simplicial 3-spheres
it would therefore be
sufficient to prove the following:
If $M$ is obtained from $M'$ by identifications
of boundary triangles in such a way that any pair of identified
triangles sits within  the same 2-sphere in $\pa M'$ and there is
some 2-sphere $\pa M'$ in which no two neighbouring triangles are
identified, then $M$ is not homeomorphic to $S^3$.  We shall in the next
section  show
that this does not hold by exhibiting an example of a simplicial
manifold $M'$ homeomorphic to a 3-ball and a pairwise
identification of the
triangles in its boundary such that no triangle is identified to a
neighbour but nevertheless the resulting manifold is simply connected
and consequently homeomorphic
to $S^3$ provided the Poincar\'{e} conjecture holds.
This does not prove the impossibility of a local construction for this
particular manifold, but certainly implies that the ordering of
identifications in the gluing process must be chosen with care if one is
to have a local construction.

\section{A counterexample}


Let us now explain the example mentioned above.
It is a little complicated so we shall explain it in a few
steps with the aid of diagrams.
First take two tetrahedra and glue them together along one triangle.
The surface of the resulting ball
is a triangulation of $S^2$ consisting of 6 triangles.
Now cut this triangulation open along one of the boundary links which
joins a vertex of degree 3 with a vertex of degree 4.
Glue to this surface a pair of triangles that are glued to each other
along two links.  Now we have a triangulation $T_1$ of $S^2$ consisting of 8
triangles.


Next cut the triangulation $T_1$
open along the links $a$, $b$ and $c$ as indicated in Fig. 2.  In this
way we obtain a two-dimensional triangulation $T_2$ with the topology of a
disc and a boundary consisting of 6 links $l_1,l_2,l_3,l_4,l_5$ and
$l_6$.  We let $l_1$ and $l_2$ be the two links that arise when we cut along
$a$, $l_3$ and $l_4$
correspond in the same way to $b$ and $l_5$ and $l_6$ correspond to $c$.
In Fig. 3 we
have drawn the dual diagram of this triangulation where a triangle is
denoted by a dot and two dots are connected by a link if and only if the
corresponding triangles share a link.  The boundary links correspond to
external legs on the dual diagram.
We label the triangles $A$, $B$, $C$, $D$, $E$, $F$, $G$ and $H$  as
indicated in Fig. 3.

Next take another pair of tetrahedra and carry out exactly the same
operations,
obtaining another triangulation $T_2'$ of the disc
with boundary links $l_1',\ldots ,l_6'$ where
$l_i'$ has the same position on the new triangulation as $l_i$ on the
first one and in the same fashion we label the triangles $A'$, $B'$ etc.

Now glue the triangulations $T_2$ and $T_2'$
together along their boundaries so that $l_i'$
is identified with $l_i$, $i=1,\ldots ,6$.
Then we obtain a triangulation $T_3$ of $S^2$
consisting of 16 triangles, see Fig. 4.
This triangulation can be extended to a triangulation of the
three-dimensional ball, e.g. by placing one vertex in the interior of the ball
and connecting this vertex to all the boundary vertices by links.
Now identify the triangles in the boundary in the
following way:
\beq
A=C,~~~B=D,~~~F=G,~~~H=E\label{id}
\eeq
and correspondingly for the primed triangles in $T_3$.  Of course two
triangles can be identified in 3 ways, but the identification is
uniquely determined by specifying two boundary links, one in each triangle,
that are
identified.  For $A$ and $C$ we identify $l_1$ with $l_2$, for $B$ and
$D$ we identify $l_3$ with $l_4$ and for $F$ and $G$ we identify $l_5$
with $l_6$.  After these identifications have been carried out there is
only one way to identify $H$ with $E$ because their boundary links have
already been identified.  The same method is used in the identification
of the primed triangles.

It is
clear from Fig. 4 that no two triangles that share a link have been
identified.  It is a straightforward counting problem
to calculate the Euler characteristic
of the complex $K$ that arises from the identifications \rf{id} and
their primed counterpart.  We
find
\beq
\chi (K)=1,
\eeq
and hence the 3-dimensional simplicial complex we have constructed
is a manifold.

We have drawn a picture of one half of the complex
$K$ in Fig. 5, i.e. the half that is made up of the unprimed triangles.
Let us call this half $\bar{K}$.  In order to see that $K$ is simply connected
one can argue as follows: Let $\delta$ be a curve in $\bar{K}$
such that the endpoints of $\delta$ lie in the boundary of $\bar{K}$,
where by the boundary of $\bar{K}$ we mean the links where $\bar{K}$
meets the other half of the complex $K$, see Fig. 5.
It is easy to see that there is a homotopy in $\bar{K}$, which keeps
endpoints fixed, from $\delta$ to another
curve $\delta '$ which lies entirely in the
boundary of $\bar{K}$.  It follows that any closed curve
in $K$ is homotopic to a closed curve in the boundary of $\bar{K}$.
The boundary of
$\bar{K}$ is contractible so $K$ is simply connected and therefore the
manifold is also simply connected.
  This completes the
discussion of the counterexample.

We remark that in
the above example we allowed two triangles to be glued along two
links. The example can easily be generalized  so that no pair of
triangles shares more than one link, by subdividing the triangles $G$ and
$H$ and their primed counterparts.

\section{Discussion}
In this note we have discussed a possible strategy to prove an
exponential bound on the number of combinatorially distinct simplicial
3-spheres as a function of volume ( = number of 3-simplexes).  We
believe that this method may be
useful for future investigations.
Partial results towards a proof have been obtained and the principal
obstacle to completing the proof has been described.

We would like to mention that
a proof of the local constructibility of all simplicial 3-spheres will,
as a consequence of Theorem 2, yield an an algorithm for constructing
all simplicial 3-spheres with a given number of tetrahedra  and hence
imply that $S^3$ is algorithmically recognizable.  The algebraic
recognizability of $S^3$
was proven recently \cite{alg1} and
can perhaps be viewed as an indication that the strategy outlined
here can be implemented.
In 5 and more dimensions it is known that spheres are not
algorithmically recognizable \cite{alg}
and there may be problems in 4 dimensions
as well \cite{benav}.  It is therefore likely that a different method will be
 needed to establish an exponential bound on $\cN (N)$ in 4 and more dimensions
if it holds in these dimensions at all.

Finally, it should be noted that proving the local constructibility
of all simply connected
simplicial 3-manifolds is a far more ambitious
project than proving this for manifolds with the topology of $S^3$.
By Theorem 2 such a result would imply the Poincar\'{e} conjecture.

\bigskip
\noindent
{\bf Acknowledgement.} We would like to thank Jan Ambj\o rn for many
discussions.  T. J. is grateful for hospitality at Institut Mittag Leffler
while this paper was completed.
This research was supported in part by a NATO science collaboration grant.

\newpage

\noindent
{\bf Figure caption.}

\bigskip\medskip

\noindent
{\bf Fig. 1.} After identifying the triangles $A$ and $A'$ the triangles
$B$ and $D$ become nearest neighbours and the same applies to the triangles
$C$ and $E$.  They may therefore be identified once
$A$ and $A'$ have been glued together.

\medskip
\noindent
{\bf Fig. 2.} The triangulation $T_1$.  We cut it open along the links marked
$a$, $b$ and $c$.


\medskip
\noindent
{\bf Fig. 3} The dual graph of the triangulation
$T_2$.  Vertices in the dual graph
correspond to triangles in the triangulation
and external legs correspond to boundary links.

\medskip
\noindent
{\bf Fig. 4.} The dual graph corresponding to the triangulation $T_3$.

\medskip
\noindent
{\bf Fig. 5.} One half of the complex $K$.  This subcomplex is made up of the
4 triangles that correspond to the original 8 unprimed triangles after they
have been identified pairwise.
There is another identical subcomplex corresponding to the primed triangles
which is glued to this half along the links $\alpha$, $\beta$ and $\gamma$.
The shaded triangle in the Figure
lies on top of another triangle and is glued to the
triangles below along two of its boundary links.

\end{document}